# QANet: Tensor Decomposition Approach for Query-based Anomaly Detection in Heterogeneous Information Networks

Vahid Ranjbar, Mostafa Salehi, Pegah Jandaghi, and Mahdi Jalili, Senior Member, *IEEE*

**Abstract**— Complex networks have now become integral parts of modern information infrastructures. This paper proposes a user-centric method for detecting anomalies in heterogeneous information networks, in which nodes and/or edges might be from different types. In the proposed anomaly detection method, users interact directly with the system and anomalous entities can be detected through queries. Our approach is based on tensor decomposition and clustering methods. We also propose a network generation model to construct synthetic heterogeneous information network to test the performance of the proposed method. The proposed anomaly detection method is compared with state-of-the-art methods in both synthetic and real-world networks. Experimental results show that the proposed tensor-based method considerably outperforms the existing anomaly detection methods.

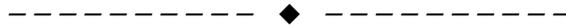

## 1 INTRODUCTION

M ANY real systems usually consist of interactions between various components and entities and can be modeled as networked structures [1]. Examples include social activities of humans, ecological systems, communication and computer networks and biological systems. Information networks are everywhere and have become a vital component of modern information infrastructure. In recent years, analysis of information networks has attracted scholars across disciplines including computer science, social sciences, mathematics and physics [2]. Modeling real-world data as information networks is a new tool that can often provide richer information as compared to traditional modeling techniques such as multidimensional modeling [3]. Representing data as information networks makes it possible to model relationships between entities. In some systems the nodes and/or edges are not from the same type and various types might coexist in the system; such systems are often modeled by heterogeneous (or multilayer) networks [4], [5], [6], [7].

One of the major challenges in the analysis of information networks is to discover anomaly and abnormal components. Anomaly detection is a branch of data mining that is associated with discovery of abnormal occurrences in the data. It has many applications in areas such as security, finance, biology, healthcare, and law enforcement [8]. For example, online social media detects spam reviews by finding unusual patterns [9]. Banks often find fraudulent activities by examining unusual transaction patterns [10]. Network intrusion detection methods detect potential network attacks by comparing traffic signatures with incoming traffic and finding unusual patterns in incoming traffic [11]. Abnormalities can be a node (entity), an edge (connection) or a subnet (a group of entities) that should not exist in the network but exist. Types of attributes and features associated with nodes and edges that are used to detect abnormalities, can be of any kind in relation to the entities or relationship between them.

So far, anomaly detection methods have mainly focused on homogeneous information networks and unstructured multidimensional data [12], [13]. Anomaly detection in heterogeneous information networks is a challenging task mainly due to specific characteristics of such networks [14]. Most of the methods developed originally for homogeneous networks do not work in the case of heterogeneous one. Often, one would like to find anomalies in certain type of nodes/edges in heterogeneous networks. For example, in bibliographic networks in which nodes can be authors, papers or venues, one might want to find author(s) that are the most different (abnormal) with others in the way they publish papers (i.e. the topic of papers and/or venues published). In this example, the difference in behavior should be a particular author chosen as reference. Such anomaly detection is referred to as query-based anomaly detection method [14].

This paper introduces a novel anomaly detection method based on tensor decomposition named QANet. In the proposed method, various meta-paths each node has with others are considered as properties of that node. Then, features are extracted using tensor decomposition, and clustering techniques are used to detect anomalies.

- V. Ranjbar and M. Salehi are with the Faculty of New Sciences and Technologies, University of Tehran, Tehran, Iran and also with the School of Computer Science, Institute for Research in Fundamental Science (IPM), P.o.Box 19395-5746, Tehran, Iran. E-mail: vranjbar@ut.ac.ir; mostafa_salehi@ut.ac.ir.
- P. Jandaghi is with the department of Computer Science, University of Southern California, California, USA, E-mail: jandaghi@usc.edu.
- M. Jalili is with the School of Engineering, RMIT University, Melbourne, Australia. E-mail: mahdi.jalili@rmit.edu.au.

We use synthetic and real datasets to evaluate performance of QANet. Specific contributions of our manuscript are as follows:

- We provide a user-centric anomaly detection approach that uses tensors to store meta-paths in heterogeneous information networks and also uses tensor decomposition techniques to extract nodal features from a tensor.
- We introduce a network generation model and an anomaly injection method to construct synthetic heterogeneous information networks to test the performance of the proposed anomaly detection method.
- We create queries from two real-world datasets including IMDB and DBLP as well as the constructed synthetic dataset and compare the proposed method with state-of-the-art methods.

The rest of the paper is organized as follows. Section 2 presents the preliminaries including a number of definitions and the problem statement. We discuss the related work in section 3. We discuss the tensor decomposition model and describe our approach for the ranking candidate set of user query in section 4. A set of comprehensive experiments is performed to evaluate the effectiveness of QANet in section 5. Section 6 draws the conclusions.

## 2 PRELIMINARIES

This section provides some preliminaries including a number of definitions required to formally state the problem of query-based anomaly detection in heterogeneous information networks.

**DEFINITION 1.** (HETEROGENEOUS INFORMATION NETWORK) [2]. A heterogeneous information network consists of multi-type entities that can have different type relationships between them, which are defined by a directed graph. Without loss of generality, the information network $G$ can be defined as $G = (V, E, F_1, F_2)$, where $V$ is the set of nodes and $E$ is set of edges. Function $F_1: V \to A$ is a function that maps each node to its type, where $A = \{A_1, A_2, ..., A_T\}$ is the set of node types and $T$ is the number of node types. Each node $v \in V$ maps to a particular type in entity type set $A: F_1(v) \in A$. Function $F_2: E \to R$ is also a function that maps each edge to its type from set $R = \{R_1, R_2, ..., R_l\}$ where $l$ is number of edge types. Each edge $e \in E$ is mapped to a special type in edge type set $R: F_2(e) \in R$. Fig. 1(a) shows a heterogeneous information network on bibliographic data. This network includes three types of node: Papers (P), Authors (A) and Venues (V). Each paper has link to a set of authors and a venue where these links belong to a set of link types.

To better understand the type of entities and relationship between them in a heterogeneous information network, it is useful to have a meta level (i.e. schema-level) description of the network.

**DEFINITION 2.** (NETWORK SCHEMA) [2]. The network schema, denoted as $T_G = (A, R)$, is a meta template for an information network $G = (V, E, F_1, F_2)$ with the node type mapping $F_1: V \to A$ and the edge type mapping $F_2: E \to R$, which is a directed graph with nodes types in $A$ and multi-type relationships from $R$. Fig. 1(b) shows network schema for bibliographic network.

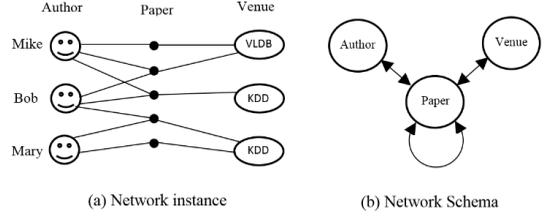

Fig. 1. (a) An instance of heterogeneous information network for bibliographic network, and (b) bibliographic network schema.

**DEFINITION 3.** (META-PATH) [2], [39] A meta path $\mathcal{P}$ is a path defined on a schema $T_G = (A, R)$, and is denoted in the form of $A_1 \xrightarrow{R_1} A_2 \xrightarrow{R_2} ... \xrightarrow{R_l} A_{l+1}$, with length of $l$. For simplicity, we can also use node types to denote the meta path if there are no multiple relation types between the same pair of node types: $\mathcal{P} = (A_1 A_2 ... A_{l+1})$. APV and APA are meta-paths for heterogeneous information network in Fig. 1.

A meta path $\mathcal{P} = (A_1 A_2 ... A_l)$ can be reversed where the reversed path is denoted as $\mathcal{P}^{-1} = (A_l A_{l-1} ... A_1)$. If $\mathcal{P}$ is equal to $\mathcal{P}^{-1}$, then $\mathcal{P}$ is a symmetric meta-path. For example, APA and APVPA are symmetric meta-paths.

**DEFINITION 4.** (META-PATH INSTANTIATION) [39]. If for each $a_i \in V$ we have $F_1(a_i) = A_i$ and for each edge $e_i(a_i a_{i+1})$ that belong to relation $R_i$ in meta-path $\mathcal{P} = (A_i A_{i+1} ... A_l)$, the path $p(a_i a_{i+1} ... a_l)$ is denoted as meta-path instance for $\mathcal{P}$. It can be denoted as $p \in \mathcal{P}$. Mike-paper1-VLDB is a meta-path instance for APV meta-path in the network of Fig. 1.

**DEFINITION 5.** (ANOMALY QUERY). An anomaly query Q is denoted by $Q = (q_c, q_R)$ where $q_c$ is a query on network entities. $S_C \subset V$ is the output indicating the outliers, known as the candidate set. $q_R$ is also a query on network entities whose output is $S_R \subset V$ serving as the reference of normal nodes. The types of referenced and candidate sets are the same. Candidate entities can be a separate set or sub-set of the reference sets.

**DEFINITION 6.** (ANOMALY SCORE). The degree of structural difference between a node and those in the reference set is the anomaly score ($\Omega_{QANet}$) of that node relative to the reference set. The structural difference means the difference in the formation of node relationship with others.

## 3 RELATED WORKS

Various methods have been proposed for anomaly detection which can be used in certain applications. Classification methods [10], [15], [16], [17], [18] require labeled data and usually provide a label for test data, which are not appropriate for applications that require a rating for abnormality. Another group of methods are based on clustering [19], [20], [21], for which efficiency is highly dependent on the clustering algorithm used. The computational complexity of these methods is challenging especially for large-scale feature set. In some other works, the nearest neighbor methods are used to detect abnormalities [22], [23], [24], [25], [26]. These methods are not suitable for datasets in which anomalies are very close to natural points, or for those in which natural data are far apart. Also, effectiveness of these methods is highly dependent on the specific distance measure used in them. Another category of anomaly detection methods are based on statistical approaches [27], [28], [29], [30], [31], [32]. The assumption of these approaches is that the data is generated from a certain distribution. However, such an assumption might not be valid in many cases, and even when the assumption is correct, it is often difficult to find the right distribution. Some other works have used information theory methods to detect abnormalities, which are more suitable for sequenced and timed data [33], [34], [35]. In [36] Noble and Cook introduced two techniques for graph-based anomaly detection based on Minimum Description Length.

Regarding the input data, anomaly detection can be divided into two categories: structured (or graph-based) and unstructured multidimensional data. In the first category, one tries to model the dependencies within the data using graphs, while in the other category the data is considered in a multidimensional space, regardless of dependencies within them. Graph-based methods can be on either homogeneous or heterogeneous graphs. Most of the previous works in anomaly detection has been on homogeneous networks or unstructured multidimensional data. In homogeneous networks, all nodes and relationships between them are of the same type. However, not all real-world networks are homogeneous. Some real systems are composed of heterogeneous types of nodes and/or edges.

There are in general two anomaly detection approaches in heterogeneous networks: approaches based on community distribution and those based on query. In community distribution approach, instead of considering the entire heterogeneous network to find possible abnormalities, distribution of nodes in communities are used. In query-based approaches, users create different queries that determine the type of anomaly and its range. For example, one might choose a particular node from a certain type and identify the most different nodes with the chosen node, as abnormal nodes in the network. Query-based approaches allow the users to interact directly with the system. The first work in the field of query-based anomaly detection was proposed by Gupta et al. in 2014 [37]. Their method considers malformation of each edge, detects anomalous groups of nodes based on a user query and returns subnets of the original network. Zhang et al. [38] provided a method for detecting anomalies based on user query to find abnormal subnets. In their proposed method, the users receive a list of abnormal subnets by defining a query. However, they did not consider the attribute of each entity, and framed the method by considering only the structural features of the network. An efficient method was proposed by Kuck et al. [14], where a formal language for queries was presented and an anomaly measure was proposed based on the network structure and existing meta-paths between the nodes. They defined an outlierness measure named as NetOut. In a heterogeneous network G and for a given query Q and for any $v_i \epsilon S_c$, the outlierness can be measured by:

$$\Omega_{NetOut}(v_i; Q) = \sum_{v_j \epsilon S_r} \frac{|\pi \rho_{sym}(v_i, v_j)|}{|\pi \rho_{sym}(v_i, v_i)|} \quad (1)$$

where smaller $\Omega$ values correspond to greater likelihood of being an outlier and $\pi \rho_{sym}(v_i, v_j)$ is the number of path instantiations of $\rho_{sym}$ (a symmetric meta-path) between two nodes $v_i$ and $v_j$. They also used PathSim and cosine similarity that were introduced in the literature to compare with their work [39]. PathSim measure between two nodes $v_i$ and $v_j$ follows a meta-path $\rho$ in a heterogeneous information network and is defined as,

$$PathSim_{\rho_{sym}}(v_i, v_j) = \frac{|\pi \rho_{sym}(v_i, v_j)|}{(|\pi \rho_{sym}(v_i, v_i)| + |\pi \rho_{sym}(v_j, v_j)|)/2} \quad (2)$$

For comparison, [14] defined:

$$\Omega_{PathSim}(v_i; Q) = \sum_{v_j \epsilon S_r} PathSim_{\rho_{sym}}(v_i, v_j) \quad (3)$$

[14] also defined a comparable version of NetOut using cosine similarity, as:

$$\Omega_{CosSim}(v_i; Q) = \sum_{v_j \epsilon S_r} \frac{\sigma \rho(v_i).\sigma \rho(v_j)}{||\sigma \rho(v_i)||_2 \times ||\sigma \rho(v_j)||_2} \quad (4)$$

where $\sigma \rho(v_i) = [|\pi \rho(v_i, v_1)|, ..., |\pi \rho(v_i, v_n)|]$ is the neighbor vector function.

These methods are state-of-the-art in this field and we compare the performance of the proposed method with them.

In proposed method, we use tensor decompositions for the anomaly detection task. The use of tensors for large-dimensional data has been of great interest in recent years [40], [41]. [42] proposed a network analysis system using tensor decomposition in order to detect malicious patterns over time. Akoglu and Faloutsos [43] proposed a tensor-based algorithm that operates on a time-varying homogeneous network and identifies anomalous points in time at which many nodes change their behavior in a way it deviates from the norm. [44] introduced a handy tool to automatically detect and visualize novel subgraph patterns within a local community of nodes in a heterogeneous network. Papalexakis et al. [45] proposed a method based on tensor decomposition for spotting anomalies in the check-in behavior of users. Koutra et al. [46] proposed a method for detection of anomalies, rare events and changes in behaviors using tensors. Although there are a number of anomaly detection methods based on tensors, query-based anomaly detection using tensor has not yet been introduced in the literature.

# 4 PROPOSED QUERY-BASED ANOMALY DETECTION APPROACH (QANET)

An outlier detection algorithm should return outliers as a subset of the candidate set, i.e. $\Omega \subset S_c$, that are considerably different from nodes in $S_R$. Let us first formally define the query-based anomaly detection problem in heterogeneous networks.

**DEFINITION 7**. (QUERY-BASED ANOMALY DETECTION PROBLEM). Let us consider the heterogeneous information network $G = (V, E, F_1, F_2)$ with node type mapping function $F_1: V \rightarrow A$ and edge type mapping function $F_2: E \rightarrow R$. Given $S_C$ as a set of candidate nodes and $S_R$ as a set of reference nodes, the problem is to return a sorted list of candidate nodes based on anomaly score, relative to nodes in the reference set. It is worth mentioning that the type of nodes in both candidate and reference sets must be the same.

In this paper, we first use tensor decomposition technique to reduce the feature set, as there are often many nodes and large number of features for each node. Then, we use clustering methods to determine the anomaly score of each node by calculating its distance with the cluster centers. Heterogeneous networks have many dimensions and representing them as a form of matrices, tables, or vectors often lead to information loss. Tensors allow us to store data in more than two dimensions, thus making it possible to store more information from the network as compared to traditional ways of network representation. Furthermore, tensor decomposition methods can be effectively used to reduce the dimensions of the feature set. One of the main disadvantages of the previous methods in this field is their high computational complexity. Tensor-based methods on the other hand are computationally effective as there are various infrastructures to implement them on distributed systems. For example, in [47] several distributed methods of decomposing tensors have been implemented on Hadoop framework.

**DEFINITION 8.** (TENSOR). An n-way (or mode) tensor is essentially a structure that is indexed by n variables. More formally, A tensor is represented by an array of $X \in \mathbf{R}^{I_1 \times I_2 \times \dots \times I_N}$.

There are a number of methods for decomposing tensors. A simple, interpretable and basic method is PARAFAC decomposition [48]. A PARAFAC model decomposes a 3-way tensor $X^{I \times J \times K}$ to trilinear components. The result is given by three loading matrices $A^{I \times F}$, $B^{J \times F}$, and $C^{K \times F}$ with elements $a_{if}$, $b_{jf}$ and $c_{kf}$ where $F$ is the number of components. The model is found to minimize the sum of squares of the residuals, $E$ in the model, where $E^{I \times J \times K}$ is a three way array of residuals:

$$X = \sum_{f=1}^{F} a_f \circ b_f \circ c_f + E \qquad (5)$$

The above relation is graphically shown in Fig. 2, for two components (F = 2). $a_f$, $b_f$ and $c_f$ are columns of the matrices $A = [a_{ij}]$, $B = [b_{ij}]$, and $C = [c_{ij}]$, respectively. The multiplications of $a_f$, $b_f$, and $c_f$ are defined as follows:

$$\left[a_f \circ b_f \circ c_f\right]_{ijk} = a_{if} b_{jf} c_{kf} \qquad (6)$$

Factors are estimated simultaneously using alternating least squares (ALS) method [49], which indeed assumes that two of the matrices are constant and the third one is estimated.

To represent a heterogeneous network by a tensor, we use the concept of meta-path. Meta-paths can indicate similarity/proximity between nodes in the network, and thus can be an important feature for anomaly detection. We define a 3-way tensor X with dimensions $N \times N \times K$ for $G$, where $N = |V|$ and $K$ is equal to the number of meta-paths with length less than 2. $X_{ijk}$ is equal to the number of instances of the kth meta-path from the list of all meta-paths extracted from the network schema with length less than 2 between nodes $v_i$ and $v_j$. $X_{ijk}$ indicates relationship between nodes $v_i$ and $v_j$ relative to the kth meta-path, which is considered as a feature for node $v_i$. QANet method constructs N×K features for each node based on meta-paths. This is often a huge number and traditional clustering algorithms such as k-mean cannot be used for that. Here we use tensor decomposition method to reduce dimension of the feature space, and thus making it possible to apply traditional clustering algorithms.

We need to compute the abnormality score for the candidate nodes relative to the reference set given by the query. Therefore, first a feature reduction model is created using tensor decomposition for the reference set. To this end, we separate a part of the tensor X that contains the features of the reference set. Let us call it $X_R$ whose dimensions are $N_R \times N \times K$, where $N_R$ is the number of nodes in the reference set. Regarding the PARAFAC decomposition method, we can obtain three matrices A, B, and C for $X_R$. Fig. 3, shows this tensor decomposition. Matrix $A$ contains the main factors of the reference set. These factors can be considered as new features for each of the nodes in the reference set. Matrix $B$ and $C$ can be used in the next step to obtain features of the candidate nodes. To obtain the features for the nodes in the candidate set, it is sufficient to use (7) to calculate $A^*$ matrix using the matrices $B$ and $C$ obtained in the previous step and $X_C$, which is defined as the $X_R$ for the candidate set nodes, as

$$A^* = X_{C_{(1)}}(C \odot B)(C^T C * B^T B)^{-1} \qquad (7)$$

where $\odot$ and $*$ are Khatri-rao and Hadamard product, respectively, and $X_{C_{(1)}}$ is mode-1 matrixization of $X_C$. After computing $A^*$, the properties of the candidate nodes are obtained using the reference nodes model. Finally, $A$ and $A^*$ are fed into the clustering algorithm as inputs (Fig. 4).

We use the K-means clustering method for the clustering phase. K-means method takes a set of observations and partitions them into k ($\leq$ n) sets so as to minimize the within-cluster sum of squares. In the clustering phase, the reference nodes are first clustered into k clusters according to the features of matrix A. Then, based on the obtained cluster centers ($C_1, \dots, C_k$), the anomaly score of each candidate node is obtained as the distance (Euclidean distance) from the center of the nearest cluster. This allows us to sort the candidate nodes based on their anomaly score and determine their final rankings:

$$\Omega_{QANet}(v_i \in S_C) = \min_{1 \le j \le k} d(A_i^*, C_j) \qquad (8)$$

where $A_i^*$ and $C_j$ are two points in Euclidean F-space and $d(A_i^*, C_j)$ is the distance between them and it is given by:

$$d(A_i^*, C_j) = \sqrt{\left(A_{i1}^* - C_{j1}\right)^2 + \dots + \left(A_{if}^* - C_{jf}\right)^2} \qquad (9)$$

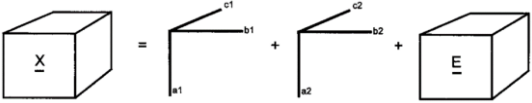

Fig. 2. A graphical representation of a PARAFAC model of tensor X [47].

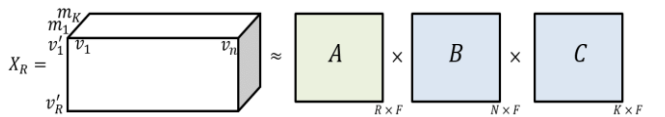

Fig. 3. Tensor decomposition for reference tensor $X_R$. We can obtain three matrices A, B, and C using PARAFAC decomposition method.

Note that QANet is based on calculating similarities using meta-paths. Homogeneous networks are a special type of heterogeneous networks that only have one type of nodes and edges. Meta-path can not be defined in these networks, and one can only define simple paths of different lengths. In the case of homogeneous networks, QANet considers similarity of nodes (e.g. common neighborhood) and identifies the anomalies based on that.

Let us consider a simple example to better understand the mechanism of QANet. TABLE 1 shows papers published by several authors, where the columns in the table represent the number of articles published by each author at various conferences. The question we want to answer is to consider an author as a reference author and rank others against the reference author on the basis of their anomaly score. As can be seen in the TABLE 1, the reference author has authored 22 papers; 10 papers published in VLDB, 10 papers in KDD, and one paper in STOC and SIGGRAPH.

TABLE 2 shows the results obtained from QANet and three other methods including CosSim, PathSim and NetOut, as state-of-the-art methods in query-based anomaly detection. To compare the performance of QANet with others, we obtained the distance measures as one minus the similarity measure. As shown in TABLE 2, all methods show Sarah exact to the reference author. In contrast to Lucy, Rob is more abnormal because Rob has published most of his papers at the conferences where the reference author has the least activity. As the result shows, in QANet method, the top anomaly is for Joe, as Joe is different to reference author both in terms of the number of papers and participated conferences. It is also seen that PathSim and CosSim also classified Joe as an anomalous author, while NetOut measure returns Joe like Sarah as a normal author. Mikel is also like Joe, but Mikel has a paper at KDD, which is one of the major conferences for the reference author, and QANet is well responding for this difference. Between Mikel and Emma, Emma has less maladaptation than Mikel, as the number of Emma's papers in the same conference as the reference author is higher than Mikel. This is only correctly captured by QANet and not by others.

## 5 RESULT

In this section we first describe the synthetic and real dataset used in this work, and then introduce the evaluation metric. We compare the performance of QANet and state-of-the-art methods in efficiently detecting anomalies. We also discuss the time complexity of QANet.

### 5.1. Data Sets

We used both synthetic and real datasets to evaluate performance of the proposed method.

*Synthetic data*

We use a similar way to the method presented in [32] to generate synthetic heterogeneous networks (Algorithm 1). We first consider $n$ nodes in $\mathcal{V}$, then assign a color to each node using the function $\psi: \mathcal{V} \rightarrow \{1, 2, \ldots, C\}$, where $C$ is

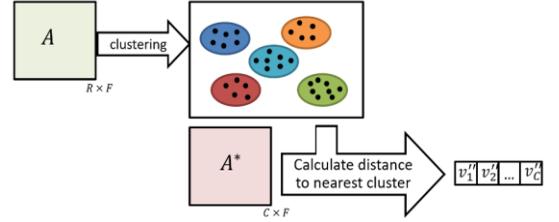

Fig. 4. Clustering the nodes in reference set with K-means clustering method and sorting the candidate nodes based on distance from the center of the nearest cluster.

TABLE 1
A TOY EXAMPLE OF BIBLIOGRAPHY NETWORK.

| NAME | VLDB | KDD | STOC | SIGGRAPH |
|---|---|---|---|---|
| Reference Author | 10 | 10 | 1 | 1 |
| Sarah | 10 | 10 | 1 | 1 |
| Rob | 0 | 1 | 20 | 20 |
| Lucy | 0 | 5 | 10 | 10 |
| Joe | 0 | 0 | 0 | 1 |
| Mikel | 0 | 1 | 0 | 0 |
| Emma | 0 | 0 | 0 | 30 |

*Taken from an example in [14] with the last row added to it.*

TABLE 2
RESULT OF ANOMALY DETECTION METHODS FOR TOY EXAMPLE PROVIDED IN TABLE 1.

| NAME | NetOut | | PathSim | | CosSim | | QANet | |
|---|---|---|---|---|---|---|---|---|
| | Score | Rank | Score | rank | Score | rank | Score | rank |
| Sarah | 0 | 6 | 0 | 6 | 0 | 6 | 0 | 6 |
| Rob | 0.9376 | 2 | 0.9 | 4 | 0.8757 | 3 | 0.81 | 4 |
| Lucy | 0.6889 | 3 | 0.6721 | 5 | 0.6717 | 4 | 0.73 | 5 |
| Joe | 0 | 4 | 0.9901 | 1 | 0.9296 | 2 | 1.02 | 1 |
| Mikel | 0 | 5 | 0.9014 | 3 | 0.2964 | 5 | 0.95 | 2 |
| Emma | 0.9667 | 1 | 0.9455 | 2 | 0.9296 | 1 | 0.88 | 3 |

the number of colors that is equal to the number of communities in the network. Also, for each $v_i$, a type $\phi(v_i)$ is considered. In order to create the graph edges, if the two nodes $v_i$ and $v_j$ are with the same color, an edge is placed between them with probability $p$, otherwise the link is placed with probability $p' \ll p$. The network is created as a heterogeneous network with different type of nodes, and nodes from the same color are connected more likely than those from different colors. This makes it possible to create groups (nodes with the same color) in the network that are similar to structure of nodes within the group. These groups represent the communities in the graph. In bibliography network for example, such groups can represent a range of research areas in which most of its authors are publishing in that area and in conferences/journals of the same domains.

*Anomaly injection*

We add some abnormal nodes to the generated network. These abnormal nodes need to be structurally different from the rest of the nodes. Thus, we randomly select a certain portion of nodes from each color, and create links between them and other nodes with probability $p''$ that is different from probabilities $p$ and $p'$.

**Input**: Number of nodes (N), Number of communities (C), set of node types (A), and probabilities $(p, p')$.
**Output**: Heterogeneous information network $G$
1:     consider $G = (V, E)$ where $V = E = \{\}$
2:     Insert $v_1$ to $v_N$ into $V$
3:     Using random function $\phi \colon \mathcal{V} \to A$ assign a type to each node $v_i$
4:     Using random function $\psi \colon \mathcal{V} \to \{1, \dots, C\}$ assign a color to each node $v_i$
5:     **For each** node pair $v_i$ and $v_j$
6:       **If** $v_i$ and $v_j$ have same color
7:         **Insert** edge $e(v_i, v_j)$ into $E$ with $p$ probability.
8:       **Else**
9:         **Insert** edge $e(v_i, v_j)$ into $E$ with $p'$ probability.
10:    Define the edge mapping function $\text{F} \colon E \to A \times A$, where $\forall\, e(v_s, v_d) \in E,\ F\big(e(v_s, v_d)\big) = \big(\phi(v_s), \phi(v_d)\big)$.
11:    **return** $G = (V, E, \phi, F)$

Algorithm 1. Pseudo-code for Generation of synthetic heterogeneous networks.

### Query generation

To generate queries, two sets of reference and candidate nodes should be selected, which needs to be from the same type. Suppose that $i$ and $j$ are two colors from the set of colors $\{1, 2, \dots, C\}$ where $C > 2$. The following query types are considered in this work.

1. We consider a number of random nodes in color $i$ as the reference set and a number of random nodes as the candidate set, half of which are in color $i$ and the other half in other colors.

2. A random number of nodes is considered as the reference set, half of which are in color $i$ and the other half in color $j$. We also consider a number of random nodes as the candidate set, half of which are in color $i$ or $j$ and the other half in other colors.

3. A number of random nodes in color $i$ is considered as the reference set, and a number of random nodes as the candidate set, half of which are anomalous nodes ($p'' = 0.5$) and the rest in color $i$.

4. We consider a random number of nodes as the reference set, half of which are in color $i$ and the other half in color $j$, and a number of random nodes as candidate set, half of which are anomalous nodes ($p'' = 0.5$) and the rest in color $i$ or $j$.

5. We consider a number of random nodes in color $i$ as the reference set and some random nodes as candidate set, half of which are in color $i$ and the rest are anomalous ($p'' = 0.5$) nodes or have color other than $i$.

6. A random number of nodes is considered as the reference set, half of which are in color $i$ and the other half in color $j$. We consider some random nodes as the candidate set, half of which are in color $i$ or $j$ and the rest are anomalous ($p'' = 0.5$) nodes or have color other than $i$ or $j$.

To evaluate the performance of the methods, we labeled nodes that have the same color as the reference nodes, as normal nodes and the rest of nodes as abnormal.

### Real data

**DBLP**: We employ a bibliographic dataset from ArnetMiner3 [50] to construct a heterogeneous information network. The dataset consists of 2,092,356 publications and 1,712,433 authors in the field of computer sci-ence. The heterogeneous network contains 3 types of vertices: paper, venue, and author. Possible type of edges includes paper-author (written-by), paper-venue (published in) and paper-paper (cited by).

**IMDb**: We use movie details dataset from the Internet Movie Database (IMDb)[1]. This dataset consists of 4,566,466 movies and 8,183,156 individuals in the role of actor, director or writer. Heterogeneous information network for this dataset contains 4 types of node: movie, actor, director and writer. Type of edges includes actor-movie (Acting), director-movie (Directing) and writer-movie (Writing).

## 5.2. Evaluation metric

In this paper, we use lift index [51] to evaluate QANet method and compare it with NetOut, PathSim, and Cos-Sim methods. Lift index measures the accuracy of a ranking method with respect to the ground truth label. The procedure for calculation of the lift index is as follows. A predictive model is first built based on the training data, which is then applied to the test data to give each test case a score showing the likelihood of the test case belonging to the positive class. The test cases are then ranked according to the scores in the descending order. After that, the ranked list is divided into $n$ equal segments, with the cases that are most likely to be positive in top segment and those that are least likely to be positive in bottom [51]. To this end, the nodes in the candidate set are ranked according to the anomaly score in the descending order, and the lift index LI is calculated as.

$$LI = \sum_{i=1}^{Can\_size} \frac{Can\_size - (i-1)}{Can\_size} \times label(v_i) \qquad (10)$$

where $Can\_size$ is the size of the candidate set and node $v_i$ is the ith node of the candidate set ranked according to the anomaly score and $label(v_i)$ is equal to 1 if the node $v_i$ is an anomalous node, and 0 otherwise. The higher is the value of LI for a method; the better is its performance. In the experiments the size of the candidate set is 10, where according to equation (10), LI takes the value between 80% (the best case) and 30% (the worst case).

## 5.3. Experimental results

In this section, we compare QANet with the methods presented by Kuck et al. [14] in 6 types of queries as above. To ensure the accuracy of the results, each of the experiments was performed 50 times, and the average results were reported. We assess the performance of the methods by varying different network parameters including the network size, the node types (degree of heterogeneity) and the number of communities in the network. We also examine two parameters related to the query: the size of the reference set and the query type. In all these settings, the two main parameters of QANet method, namely, the rank of the tensor decomposition and the number of clusters in the clustering method, are studied. When comparing the methods for different network parameters, we use the query type 5, which has both anomalous nodes ($p'' = 0.5$) and those with colors than the reference set.

---

[1] https://www.imdb.com/interfaces/

### 1) Network size

Fig. 5 shows the lift index of QANet as a function of the network size. For this simulation, we consider the decomposition ranks 4 and 8, and the number of clusters 1, 4, and 20. As it can be seen, the decomposition rank does not have much influence on the accuracy of the proposed method, however as the number of clusters increases, its accuracy decline, i.e. the lift index decreases. As the nodes in the reference set are from the same color (query type 5), by choosing a cluster size of 1, all nodes in the reference set are placed in one cluster, and thus improve the accuracy of the proposed method in finding the abnormal nodes.

Fig. 6 compares the accuracy of NetOut, PathSim, CosSim and QANet for varying network sizes. As it is seen, the proposed method (QANet) has considerably better accuracy than others. Also, as the network size increases, its accuracy also increases. This is due to the fact QANet is based on network structure, and as the network becomes larger, more features can be extracted for nodes, resulting in improved detection of abnormalities. Other methods have either decreased or unchanged behavior as the network size increases. These results indicate the QANet is more suitable for large-scale network than other state-of-the-art methods.

### 2) Node types

Another parameter that affects anomaly detection is node types, which is indeed an indicator of heterogeneity level in the network. Increasing the node types while the number of nodes is kept unchanged, has almost the same effect as decreasing the number of nodes. Thus, one would expect declined accuracy for increased node types. On the other hand, by increasing the node types, the number of edges in the network decreases with respect to the generation model, which makes it more difficult to distinguish the node from one another. However, as it is seen from the results, QANet is almost insensitive to node type, while other methods have rather more changes (Fig. 7).

### 3) Number of communities

Fig. 8, shows the impact of the number of communities in the network on the accuracy of the methods. In many heterogeneous networks the nodes can be divided into different groups. For example, in bibliographic networks there are different scientific fields, and each node (author, paper or conference) is in one or more of them. Our results show that by increasing the number of communities, the accuracy of QANet decreases (Fig. 8). Indeed, increasing the number of communities in the network reduces the distinction between the nodes, which reduces the accuracy of the QANet. Other methods however are not considerably impacted by the number of communities.

### 4) Size of reference set

Fig. 9, shows the accuracy of the algorithms with respect to the size of the reference set. As the size of the reference state increases, the accuracy of QANet is systematically improved, whereas other methods do not follow any specific pattern as a function of the size of the reference state.

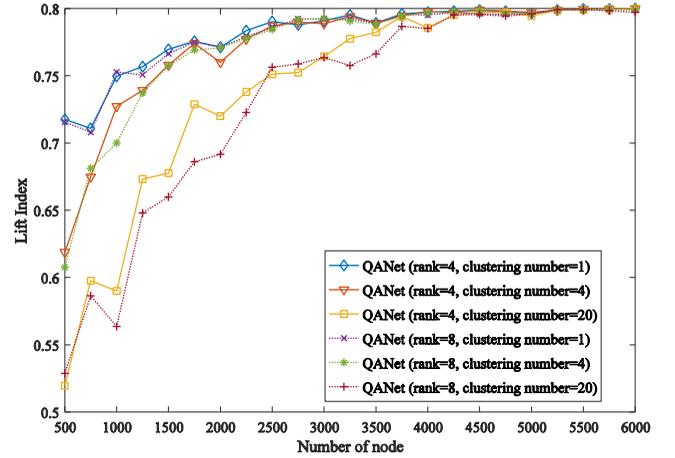

Fig. 5. The accuracy of QANet as a function of the network size in synthetic network. In this experiment, we consider the decomposition ranks 4 and 8, and the number of clusters 1, 4, and 20. Also we set the node types as 2, number of communities is 4 and query type is 5.

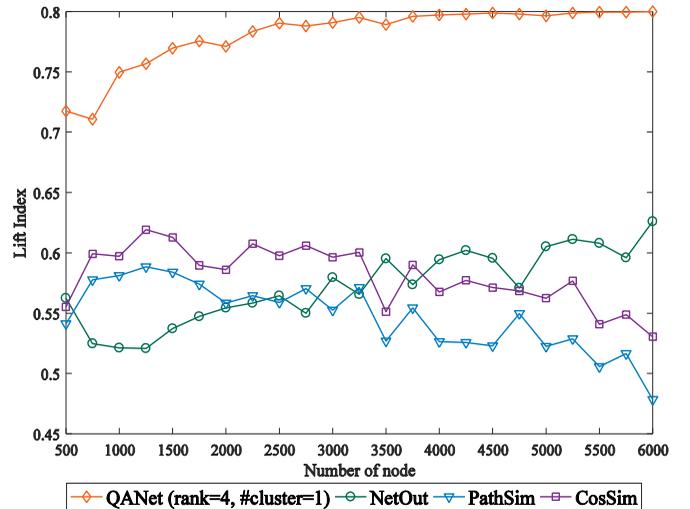

Fig. 6. Accuracy of NetOut, PathSim, CosSim and QANet as a function of network szie in synthetic networks. In this experiment, we set the node type as 2, the number of communities as 4, and the query type 5.

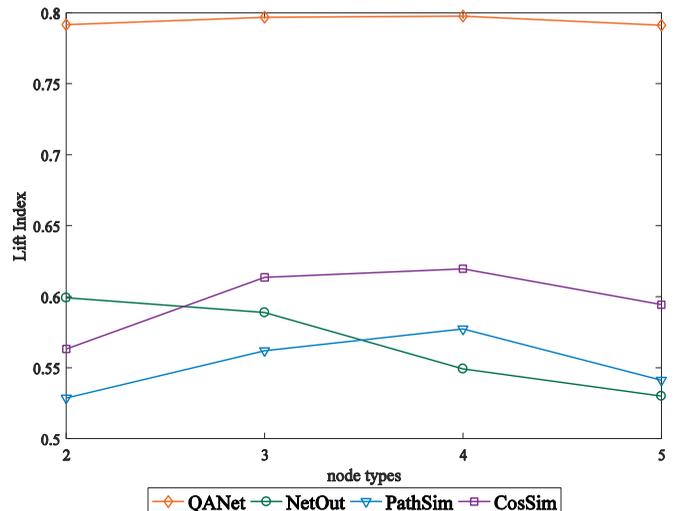

Fig. 7. Accuracy of the methods as a function of node type in synthetic networks. In this experiment, the number of nodes is 4000, the size of the reference set is 20, and there query type is 5. We set the number of cluster to 1 and the decomposition rank to 4 for QANet.

Improved performance of QANet as a function of this parameter is due to enhancing the ability of the method through reconstructing a more precise model by having a richer feature set. As in NetOut, PathSim, and CosSim methods, only a simple averaging is considered for nodes in the reference set, increasing the number of nodes in the references set might worsen the performance, which is clearly seen in the results.

### 5) Query type

Fig. 10 illustrates the accuracy of the methods for the six types of queries. As shown in the Fig. 10, QANet is the top-performer in all query types. While PathSim and CosSim show similar performance as QANet in query types 3 and 4, their accuracy is not comparable with QANet in other query types. NetOut has higher accuracy than PathSim and CosSim method in query types 1 and 2, but lower in other types. Because they can detect anomalous nodes ($p'' = 0.5$) correctly, but cannot detect nodes from other communities while NetOut cannot detect anomalous nodes. Therefore, because there are no anomalous nodes in Type 1 and 2 queries, the NetOut is better and PathSim and CosSim are better in the other queries. But as shown in the Fig. 10, the proposed method can well detect both types of abnormalities.

## 5.4. Case studies

In this section, we examine the effectiveness of QANet and NetOut on two real datasets.

### Case 1: Bibliographic Dataset

In DBLP dataset, we consider all authors who collaborate with Christos Faloutsos as candidate set and Christos Faloutsos as reference set. In this query, the reference set has only one member and the candidate set contains 426 authors. According to the definition of anomalies in the section 2, the abnormalities indicate the structural difference in the communication (i.e. different meta-paths) between any of the nodes in the candidate set and the node(s) in the reference set. 10 authors who have the highest degree of malformation based on QANet and NetOut methods are listed in Tables 3 and 4, respectively.

As seen in Table 3, the top-4 abnormal authors provided by QANet algorithm each have only one paper co-authored with the reference author. Each of these authors has published only their paper at the conference in which the reference author has published only a paper. Furthermore, this conference was not one of the major conferences of the reference author. Also, they are arranged based on the citation count that indicates the importance of their paper and their collaboration with the reference. The next four authors in Table 3 have also published only one paper, however they have a common co-author with the reference, thus strengthening their relationship with the reference. These four authors are also arranged based on the citation count that indicates the importance of their collaboration with the reference author, which also reflects the overall view of the QANet. Lei Li is ranked 9. While the publications of this author are quite similar to the top-ranked author. The venue that this author has published his paper is more important to the reference, as he has published another paper at this venue. This indicates that in contrast to Philip, Lei Li is closer to Christos Faloutsos as there are more paths between them. Finally, the last author in the ranking Table 3 has also coauthored one paper with the reference author, however this author has two collaborators who have co-authored with the reference, making him closer to the reference as compared to those top in the ranking table. These results indicate that the proposed ranking algorithm leads to reasonable outcome.

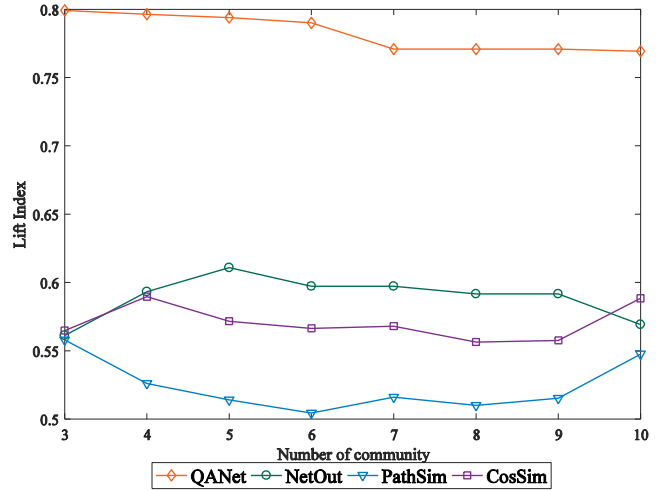

Fig. 8. Accuracy of the methods as a function of the number of communities in synthetic networks. In this experiment, the number of nodes is 4000, there are two types of nodes, the size of the reference set is 20, and there query type is 5. We set the number of cluster to 1 and the decomposition rank to 4 for QANet.

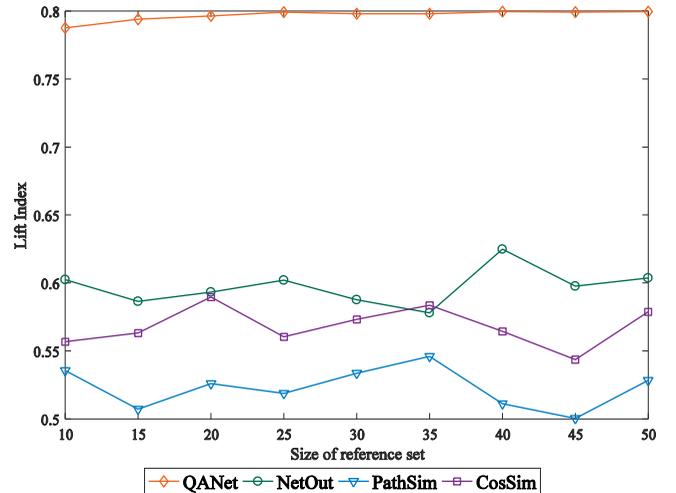

Fig. 9. Accuracy of the methods as a function of the reference size in the query for synthetic networks. In this experiment, number of nodes is 6000, node types is 2 and the number of communities is 4, and query type is 5. In QANet number of clusters is 1 and decomposition rank is 4.

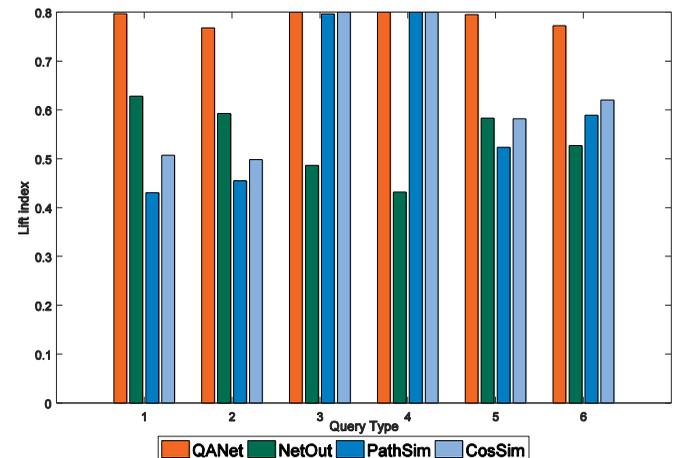

Fig. 10. Accuracy of the methods as a function of the query types described in Section 5.1 for synthetic networks. In this experiment, the number of nodes is 6000, node types is 2, the number of communities is 4 and the size of the reference set is 20. We set the number of cluster to 1 and the decomposition rank to 4 for QANet.

In order to have a better understanding on the performance of competing algorithms, Table 4 shows the top-10 abnormal authors to the reference author based on NetOut. This way the query type based upon Tables 3 and 4 are extracted is exactly the same. NetOut requires the users to specify the meta-path in addition to the reference and candidate sets. In order to have a fair comparison with QANet, we chose all meta-paths of length 2 used in QANet. As seen from the results of the ranking obtained by NetOut method, this method only works on the basis of counting meta-paths. This method ranks the authors based on their publication pattern, i.e. the number of articles and citation, as compared with the reference author. The method also takes into account joint publications with the references author. Table 5 shows the top-10 dissimilar authors to the reference author based on these two algorithms. Clearly, these two ranking algorithms result in considerably different outcome. However, as it is clearly seen in Tables 3 and 4, QANet results in more reasonable ranking outcome than NetOut, as the authors ranked by QANet have more dissimilarity with the reference author than those ranked by NetOut.

### Case 2: Internet Movie Database

In this case study, according to the definition of anomalies in section 2, we want to identify abnormal actors, among actors of "The Godfather (1972)" movie. According to the Internet Movie Database (IMDb) site, 34 of the actors are introduced as the main cast for this movie, and we consider them as the candidate and reference set. Table 6 lists the actors and their details, as well as the ranking of QANet and NetOut methods. We set the number of cluster to 1 and the decomposition rank to 3 for QANet.

In order to confirm the results of the anomaly detection methods, we show the matrices of AMA, AMAMA, AMDMA and AMWMA meta-paths for actors in the form of heatmap in Fig. 11. With regard to various meta-paths, *James Caan*, *Marlon Brando*, *Al Pacino*, *Robert Duvall* and *Diane Keaton* are more abnormal than the rest of the actors. These actors are exactly the five top-ranking actors in QANet ranking, while NetOut method ranked them 3, 13, 15, 9 and 7, respectively.

In order to better understand our method, we show the results of tensor decomposition with rank 3 and clustering in a three-dimensional space in Fig. 12. As shown in

Fig. 12, the abnormal actors are much farther away than the rest of the actors.

### 5.5. Time complexity

This section provides some analysis on the computational

**TABLE 3**
**Top-10 ranked authors by QANet**

| Ranking | Author's Name | Number of papers | Number of APA Meta-Paths with Faloutsos | Number of APA Meta-Paths with Faloutsos | Number of APA-PA Meta-Paths with Faloutsos | Number of paper's Citation | Number of APVPA Meta-Paths with Faloutsos |
|---|---|---|---|---|---|---|---|
| 1 | Philip Russell (Flip) Korn | 1 | 1 | 0 | 0 | 1 |
| 2 | George Panagopoulos | 1 | 1 | 0 | 3 | 1 |
| 3 | Ibrakim Kamel | 1 | 1 | 0 | 11 | 1 |
| 4 | Yi Rong | 1 | 1 | 0 | 21 | 1 |
| 5 | Caetano Traina Jnior | 1 | 1 | 1 | 0 | 1 |
| 6 | Robson L. F. Cordeiro | 1 | 1 | 1 | 0 | 1 |
| 7 | Yi Zhou | 1 | 1 | 1 | 6 | 1 |
| 8 | Bin Zhang | 1 | 1 | 1 | 6 | 1 |
| 9 | Lei Li | 1 | 1 | 0 | 0 | 2 |
| 10 | Wenyao Ho | 1 | 1 | 2 | 5 | 1 |

*APA indicates author-paper-author meta-path, and APAPA and APVPA stand for author-paper-author-paper-author meta-path, author-paper-venue-paper-author meta-path, respectively.*

**TABLE 4**
**Top-10 ranked authors by NetOut**

| Ranking | Author's Name | Number of papers | Number of APA Meta-Paths with Faloutsos | Number of APA Meta-Paths with Faloutsos | Number of APA-PA Meta-Paths with Faloutsos | Number of paper's Citation | Number of APVPA Meta-Paths with Faloutsos |
|---|---|---|---|---|---|---|---|
| 1 | Sebastian B. Thrun | 139 | 1 | 75 | 799 | 1 |
| 2 | Venkatesan Guruswami | 124 | 1 | 518 | 1962 | 5 |
| 3 | Arthur Toga | 84 | 1 | 470 | 478 | 1 |
| 4 | K. P. Sycara | 252 | 1 | 640 | 3757 | 11 |
| 5 | Asim Smailagic | 73 | 1 | 514 | 667 | 3 |
| 6 | N. Sadeh | 88 | 1 | 558 | 799 | 11 |
| 7 | Dan Siewiorek | 192 | 1 | 662 | 1962 | 65 |
| 8 | David A. Bader | 94 | 1 | 536 | 478 | 2 |
| 9 | Douglas W. Oard | 112 | 1 | 497 | 724 | 8 |
| 10 | M Hebert | 131 | 1 | 534 | 1943 | 7 |

**TABLE 5**
**Top-10 ranked authors based on QANet and NetOut with Christos Faloutsos as the reference set**

| | Top-10 ranked authors based on QANet | | | Top-10 ranked authors based on NetOut | |
|---|---|---|---|---|---|
| QANet Ranking | Author's Name | NetOut Ranking | QANet Ranking | Author's Name | NetOut Ranking |
| 1 | Philip Russell (Flip) Korn | 421 | 196 | Sebastian B. Thrun | 1 |
| 2 | George Panagopoulos | 420 | 197 | Venkatesan Guruswami | 2 |
| 3 | Ibrakim Kamel | 417 | 151 | Arthur Toga | 3 |
| 4 | Yi Rong | 418 | 252 | K. P. Sycara | 4 |
| 5 | Caetano Traina Jnior | 408 | 170 | Asim Smailagic | 5 |
| 6 | Robson L. F. Cordeiro | 409 | 209 | N. Sadeh | 6 |
| 7 | Yi Zhou | 406 | 245 | Dan Siewiorek | 7 |
| 8 | Bin Zhang | 407 | 216 | David A. Bader | 8 |
| 9 | Lei Li | 422 | 194 | Douglas W. Oard | 9 |
| 10 | Wenyao Ho | 395 | 224 | M Hebert | 10 |

complexity of the QANet. The QANet consists of several steps. The first step is to extract queries and prepare the reference and candidate sets for queries. The second stage involves the implementation of the tensor decomposition method and obtaining the candidate and reference matrices. The final step is to apply the clustering method to the characteristics obtained from the previous steps for the candidate and reference sets. In the first step, the query language provided by Kuck et al. [14] is used. Network tensor and meta-paths of length 2 can be pre-calculated offline. Sparse tensor is used in the implementation due to its rather low time complexity. If the number of nodes in the reference and candidate sets are $N_R$ and $N_C$, respectively, the complexity of obtaining two tensors $X_R$ and $X_C$ is equal to $O(N_C + N_R)$. In the second step for decomposing tensors and obtaining the properties of each node, according to alternating least squares (ALS) algorithm, the worst case time complexity is $O(RN_CNK + tRN_RNK)$, where $R$ is the decomposition rank (usually less than 10) and $t$ is the number of iterations of the tensor decomposition method (50 in this manuscript), and $N$ and $K$ are the number of nodes in the network and the number of meta-paths with length less than 2, respectively. $K$ is at most $T^3$, where $T$ is the node types in the network. In the final step, according to the k-means clustering method and calculation of the distance between each candidate node and the cluster centers, time complexity is equal to $O(CN_C + iCN_R)$, where $i$ is the number of clustering iteration and $C$ is the number of clusters. Finally, the time complexity of QANet is $O((N_C + N_R)N)$.

NetOut, PathSim and CosSim methods have time complexity exponential to the meta-path length. Materializing neighbor vector requires traversal of the heterogeneous network, which can be time-consuming when the specified meta-path is long or when the degree of the node of interest is high. The time complexity of these methods is $O(N_CN_RN)$ when meta-paths of length 2 are considered. It is more complicated for meta-paths of higher lengths.

### TABLE 6
The cast of "The Godfather (1972)" and their details, as well as the ranking of QANet and NetOut methods.

| Row | Actor's Name | Actor | Director | Writer | QANet Ranking | NetOut Ranking |
|---|---|---|---|---|---|---|
| 1 | Richard Conte | 105 | 0 | 1 | 13 | 4 |
| 2 | Corrado Gaipa | 34 | 0 | 0 | 23 | 12 |
| 3 | Morgana King | 46 | 0 | 0 | 17 | 27 |
| 4 | Lenny Montana | 6 | 1 | 0 | 16 | 31 |
| 5 | James Caan | 330 | 0 | 1 | 2 | 3 |
| 6 | John Cazale | 14 | 0 | 0 | 28 | 29 |
| 7 | Sterling Hayden | 84 | 1 | 0 | 32 | 11 |
| 8 | Talia Shire | 48 | 0 | 1 | 30 | 19 |
| 9 | Abe Vigoda | 150 | 0 | 0 | 10 | 1 |
| 10 | Marlon Brando | 174 | 0 | 1 | 4 | 13 |
| 11 | Rudy Bond | 38 | 0 | 0 | 22 | 17 |
| 12 | Richard Bright | 30 | 0 | 0 | 19 | 21 |
| 13 | Richard S. Castellano | 51 | 0 | 0 | 27 | 22 |
| 14 | Franco Citti | 55 | 1 | 1 | 34 | 10 |
| 15 | Salvatore Corsitto | 3 | 0 | 0 | 21 | 34 |
| 16 | Al Pacino | 130 | 2 | 4 | 5 | 15 |
| 17 | Tony Giorgio | 6 | 0 | 0 | 14 | 32 |
| 18 | Julie Gregg | 41 | 0 | 0 | 7 | 14 |
| 19 | Angelo Infanti | 72 | 0 | 0 | 15 | 5 |
| 20 | Robert Duvall | 268 | 4 | 5 | 1 | 9 |
| 21 | Al Lettieri | 14 | 1 | 0 | 31 | 24 |
| 22 | Jeannie Linero | 23 | 0 | 0 | 12 | 18 |
| 23 | Tere Livrano | 4 | 0 | 0 | 20 | 33 |
| 24 | John Marley | 108 | 0 | 0 | 29 | 8 |
| 25 | Al Martino | 62 | 0 | 0 | 33 | 23 |
| 26 | John Martino | 13 | 0 | 0 | 24 | 28 |
| 27 | Diane Keaton | 166 | 1 | 14 | 3 | 7 |
| 28 | Victor Rendina | 7 | 0 | 0 | 9 | 20 |
| 29 | Alex Rocco | 143 | 0 | 0 | 8 | 2 |
| 30 | Gianni Russo | 18 | 2 | 0 | 6 | 26 |
| 31 | Vito Scotti | 75 | 0 | 0 | 25 | 6 |
| 32 | Ardell Sheridan | 13 | 2 | 0 | 18 | 30 |
| 33 | Simonetta Stefanelli | 13 | 0 | 0 | 26 | 25 |
| 34 | Saro Urzì | 24 | 0 | 0 | 11 | 16 |

*The actor column indicates the number of movies that they act and the director column indicates the number of movies directed by them and the writer column indicates the number of movies written by them.*

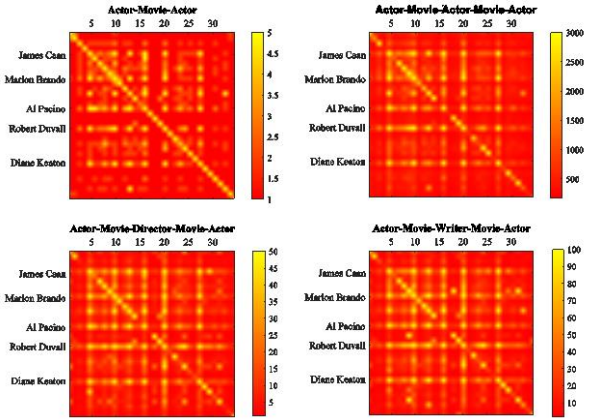

Fig. 11. Adjacency matrix of Actor-Movie-Actor, Actor-Movie-Actor-Movie-Actor, Actor-Movie-Director-Movie-Actor and Actor-Movie-Writer-Movie-Actor metapaths for actors of "The Godfather (1972)".

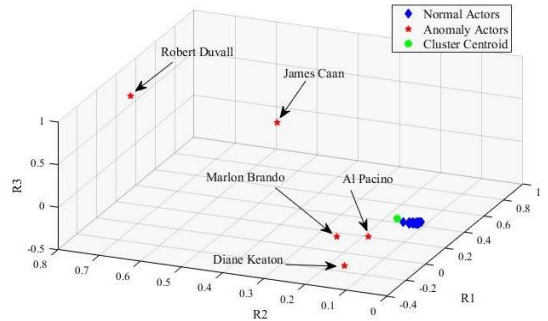

Fig. 12. Illustration of the QANet method. We set the number of cluster to 1 and the decomposition rank to 3 for QANet. R1, R2 and R3 are columns of $A^*$ respectively. Cluster centroid indicates center of cluster after clustering phase.

# 6 CONCLUSION

In this paper, we proposed a tensor decomposition method called QANet for detecting query-based anomalies in heterogeneous information networks. QANet considers all the different aspects of the communication structure and uses the PARAFAC tensor decomposition method to create a model, which is then used to rank the candidate nodes based on their abnormality (dissimilarity) with those in the reference set. Due to lacking tagged data, we introduced a model to create synthetic heterogeneous information networks, and tested effectiveness of the proposed anomaly detection algorithm for various query types. We also compared the performance of QANet with state-of-the-art algorithms including NetOut, PathSim and CosSim. The experiments showed that QANet results in better performance by providing more accurate prediction of abnormal nodes than other algorithms. We also compared the performance of QANet and NetOut on two real datasets: bibliographic network and Internet Movie Database (IMDb). The results revealed that the ranking provided by QANet is more reasonable than the one provided by NetOut. QANet outperformed NetOut. This is mainly due to the fact that in QANet all meta-paths of the candidate and reference nodes with other network nodes are considered. However, in NetOut method, only the symmetric meta-paths between the candidate and reference nodes are considered, while the relationships of these nodes with other network nodes are not considered. Indeed, QANet takes into account more global information in building the similarity metrics for the anomaly detection. Our experimental results confirm that QANet performs better. Future directions to the research work introduced here include evaluating our approach on temporal networks as well as using it for event detection in time-evolving networks.

## ACKNOWLEDGMENT

This work was supported in part by a grant from IPM (No. CS1396-4-49) and is partially supported by Iran National Science Foundation (INSF) (Grant No. 96001338). Mahdi Jalili is supported by Australian Research Council through project No DP170102303. Mostafa Salehi is the corresponding authorship for paper.

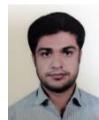
**Vahid Ranjbar.** Received the B.S. and the M.S. degree in information technology in 2011 and 2013 respectively. He is currently working toward the Ph.D. degree in the University of Tehran, Iran. His research interests include network science and data mining.

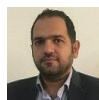
**Mostafa Salehi.** completed his PhD studies in Computer Engineering at Sharif University of Technology, Iran in 2012. On 2013, he joined University of Tehran as an assistant professor. His research interests include network science and multimedia networks.

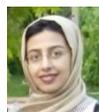
**Pegah Jandaghi.** Received the B.S. degree in Computer Engineering and Mathematical and application in 2017. he is currently working toward the MS degree in the University of Southern California, USA. Her research interests include Information Networks and data mining.

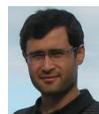
**Mehdi Jalili.** (Member 2009, SM 2016) received the PhD degree from EPFL (Swiss), in 2008. He is now a senior lecturer with the School of Engineering, RMIT University, Melbourne, Australia. His research interests are in network science, dynamical systems, data mining.